\documentclass[journal,twoside,web]{ieeecolor}
\usepackage{tmi}
\usepackage{cite}
\usepackage{amsmath,amssymb,amsfonts}
\usepackage{algorithmic}
\usepackage{graphicx}
\usepackage{textcomp}
\usepackage{mathtools}
\usepackage{algorithm}
\usepackage{algorithmic}
\usepackage{textcmds}
\usepackage{url}

\definecolor{revisedColor}{rgb}{0,0,1}

\def\BibTeX{{\rm B\kern-.05em{\sc i\kern-.025em b}\kern-.08em
    T\kern-.1667em\lower.7ex\hbox{E}\kern-.125emX}}
\begin{document}
\title{\LARGE Generative Consistency Models for Estimation of Kinetic Parametric Image Posteriors in Total-Body PET}
\author{Yun Zhao, Qinlin Gu, Georgios I. Angelis, Andrew J. Reader, Yanan Fan, Steven R. Meikle
\thanks{This work is supported by the Australian Research Council (Discovery Projects grant DP230102070).}
\thanks{Yun Zhao (yun.zhao@sydney.edu.au) is with the School of Health Sciences and the Brain and Mind Centre, the University of Sydney, Australia.}
\thanks{Qinlin Gu is with the School of Mathematics and Statistics and the Brain and Mind Centre, the University of Sydney, Australia.}
\thanks{Georgios Angelis is with Sydney Imaging, the University of Sydney, Australia.}
\thanks{Andrew J. Reader is with the School of Biomedical Engineering and Imaging Sciences, King’s College London, UK.}
\thanks{Yanan Fan is with Data61, CSIRO, Australia.}
\thanks{Steven R. Meikle (steven.meikle@sydney.edu.au) is with the School of Health Sciences, Sydney Imaging and the Brain and Mind Centre, the University of Sydney, Australia.}}

\maketitle

\begin{abstract}
Dynamic total body positron emission tomography (TB-PET) makes it feasible to measure the kinetics of the tracer in all organs of the body simultaneously which may lead to important applications in multi-organ disease and systems physiology. Since whole-body kinetics are highly heterogeneous with variable signal-to-noise ratios, parametric images should ideally comprise not only point estimates but also measures of posterior statistical uncertainty. However, standard Bayesian techniques, such as Markov chain Monte Carlo (MCMC), are computationally prohibitive at the total body scale. We introduce a generative consistency model (CM) that generates samples from the posterior distributions of the kinetic model parameters given measured time-activity curves and arterial input function. 
CM is able to collapse the hundreds of iterations required by standard diffusion models into just 3 denoising steps.
The CM was evaluated using physiologically realistic simulations and an application to a subject's dynamic [$^{18}$F]FDG TB-PET dataset analyzed with a standard single-input two-tissue compartment model.
When trained on 500,000 physiologically realistic two-tissue compartment model simulations, the CM produces similar accuracy to MCMC (median absolute percent error $< 5\%$; median K-L divergence $< 0.5$) but is more than five orders of magnitude faster. CM produces more reliable $\mathit{K_i}$ images than the Patlak method by avoiding the assumption of irreversibility, while also offering valuable information on statistical uncertainty of parameter estimates and the underlying model.
The proposed framework removes the computational barrier to routine, fully Bayesian parametric imaging in TB-PET and is readily extensible to other tracers and compartment models.
\end{abstract}


\section{Introduction}
Dynamic positron emission tomography (PET) provides time-resolved measurements of a radiotracer distribution in the body, to which compartmental models can be fit for the estimation of physiological parameters, such as the net influx rate constant, $K_i$. These quantitative parameters enable the testing of hypotheses in research studies, while in clinical studies they may provide additional discriminatory power beyond that of the semi-quantitative standardised uptake value (SUV) \cite{dias2021clinical}. Common parameter estimation techniques include nonlinear least squares and linear graphical analyses, such as the Patlak plot, which are more computationally efficient  \cite{patlak1983graphical,koeppe1985performance,hong2010kinetic,kang2020robust}.
The recent introduction of total-body and long axial field of view ($\ge1$ metre) PET scanners makes it possible to acquire dynamic data from all major organs of the body simultaneously. 
Current efforts in total-body parametric imaging have primarily utilized graphical methods like Patlak to manage the computational cost of estimating parameters over millions of voxels \cite{rahmim2019dynamic,karakatsanis2013quantitative}. 
However, these graphical approaches often suffer from bias due to simplifying assumptions (e.g., irreversibility), while standard nonlinear least squares (NLS) remains computationally prohibitive for such large volumes.
Emerging deep learning approaches have begun to address these challenges \cite{wang2021total,liang2023combining,wang2020noninvasive}, yet they predominantly focus on non-Bayesian estimation, leaving the critical aspect of uncertainty quantification largely unexplored.
Therefore, the gap is significant because total-body imaging introduces substantial kinetic heterogeneity, calling into question the appropriateness of assuming a single kinetic model for all organs and tissues. Furthermore, signal-to-noise ratios are highly variable throughout the body, making it increasingly important to compute not only point estimates of kinetic parameters but also measures of their statistical uncertainty to assess reliability.

Bayesian inference addresses this need by computing the posterior distributions, $p(\mathbf{\theta}|y)$, of the kinetic model parameters, $\mathbf{\theta}$, given a time-activity curve (TAC), $y$. Markov chain Monte Carlo (MCMC) is the reference method for generating asymptotically unbiased posteriors \cite{brooks2011handbook,gelman2013bayesian}. However, MCMC requires an analytical expression for the likelihood function based on an assumed noise distribution, which is not reliably known at the voxel level when using non-linear reconstruction methods such as expectation maximization, as well as a long burn-in phase and careful assessment of convergence.
Even when convergence is reached, MCMC samplers are slow and are not easy to parallelize. 
In dynamic total-body PET (TB-PET) studies, where millions of voxels require independent MCMC chains (a single study may contain millions of voxels), these characteristics make MCMC impractical \cite{irace2020bayesian}.
Likelihood-free alternatives, such as approximate Bayesian computation (ABC), improve the situation by bypassing the need to compute the likelihood, but they are still computationally highly inefficient \cite{sunnaaker2013approximate}.
Variational Bayesian methods offer a faster alternative, but their approximations can lead to biased parameter estimates \cite{blei2017variational}.
As a result, the application of Bayesian computation methods for posterior probability estimation in PET is largely confined to the analysis of regions of interest.

Deep generative models have recently emerged as computationally efficient alternatives to MCMC. Conditional variational auto-encoders (CVAEs), generative adversarial networks (GANs) and denoising diffusion probabilistic models (DDPMs) have all been applied to medical imaging to estimate posteriors \cite{liu2023posterior,mucke2023markov,djebra2025bayesian}.
Nonetheless, these approaches have limitations. CVAEs underestimate posterior variance \cite{blei2017variational}, GANs suffer from mode collapse \cite{salimans2016improved}, and diffusion models require hundreds to thousands of reverse steps \cite{ho2020denoising}, which scales poorly to TB-PET image volumes. Moreover, existing studies have focused mainly on ROI-based analysis of the brain \cite{djebra2025bayesian} and the application of generative models to dynamic TB-PET, with its broader range of kinetics and greater inter-organ heterogeneity, has not been demonstrated.

The present work introduces a conditional consistency model (CM) for posterior estimation in dynamic TB-PET. The CM reformulates diffusion sampling as a short schedule of denoising operations and has been shown to retain the theoretical guarantees of score-based diffusion (SBD) while collapsing the sampling path to one step \cite{song2023consistency}.
We adapt this idea to PET compartmental modelling by training a CM to learn a direct mapping from pure Gaussian noise samples to kinetic parameter samples, conditioned on both the TAC and the arterial input function (AIF). Trained on a large ensemble of physiologically realistic simulations, the proposed method produces voxel-wise posterior samples for the given kinetic model within a practical time frame, without sacrificing accuracy. We evaluate the performance of the CM by comparison with the reference method MCMC, as well as ABC and common generative deep learning approaches. We then illustrate a practical application of the CM method to a dynamic total body [\textsuperscript{18}F]FDG PET study, and we show how results from CM can be used to perform voxel-level model selection tasks to account for kinetic heterogeneity and to compute parametric images of posterior uncertainty in estimating $K_i$.

\section{Methods}
\subsection{Consistency Models}
CMs are a class of generative models that address one of the main limitations of diffusion-based methods: the lengthy iterative process required to generate samples from the posterior given an observation of noisy data. Diffusion models \cite{ho2020denoising,song2020score} typically work by progressively denoising a Gaussian noise sample, following an iterative procedure. However, this multi-step denoising process introduces a severe time‐efficiency bottleneck, as generating a single sample can require hundreds of model evaluations.
Consistency models were proposed to alleviate this issue by significantly reducing the number of steps required to generate high-quality samples, thereby accelerating the prediction process without compromising fidelity \cite{song2023consistency}. 

An extension of these models involves conditioning them on auxiliary information, denoted by $y$. Mathematically, let $f_\phi(\mathbf{\theta}_t,t,y)$ denote the denoising neural network with trainable parameters $\phi$ that can be used to predict a posterior sample from a noisy input $\mathbf{\theta}_t$ at a noise level $t$, conditioned on the auxiliary data $y$. 
Note that in this study, $\theta$ is a vector of kinetic parameters and $y$ is a concatenated vector of the TAC and the AIF.
Suppose $\mathbf{\theta}_0$ is the ground-truth target, and $\mathbf{\theta}_t=\mathbf{\theta}_0+\sigma_t z$ represents a noisy version of $\mathbf{\theta}_0$, where $\sigma_t$ is a noise schedule and $z$ is a standard Gaussian noise sample. The consistency condition states that for $0\leq t_1\leq t_2\leq1$,
\[
    f_\phi (\mathbf{\theta}_{t_1},t_1,y)=f_\phi (\mathbf{\theta}_{t_2},t_2,y)=\theta_0,
\]
ensuring that denoising at time $t_1$ yields the same result as denoising at $t_2$. The corresponding training algorithm used in this work follows the Consistency Training in the original paper \cite{song2023consistency} with some adaptations, summarized in \textbf{Algorithm} \ref{alg:consistency-training}.
\begin{algorithm}[htp]
  \caption{Consistency Training}
  \label{alg:consistency-training}
  \begin{algorithmic}[0]
    \REQUIRE dataset of training pairs $\mathcal{D} = \{(\mathbf{\theta},y)\}$, initial consistency model parameters $\phi$, learning rate $\eta$, fixed step size $N$, fixed exponential moving average decay $\mu$, distance $d(\cdot,\cdot)$\\
    $\phi^- \gets \phi$
    \REPEAT
      \STATE Sample $(\mathbf{\theta}, y) \sim \mathcal{D}$ and $n \sim \mathcal{U}\bigl[\,1,\,N\bigr]$
      \STATE Sample $z \sim \mathcal{N}(0, I)$
      \STATE \(
        \mathcal{L}(\phi, \phi^-) \;\gets\;
        d\bigl(
          f_{\phi}\bigl(\mathbf{\theta} + t_{n+1}\,z,\;t_{n+1},\,y\bigr)\,,\;
          f_{\phi^-}\bigl(\mathbf{\theta} + t_n\,z,\;t_n,\,y\bigr)
        \bigr)
      \)
      \STATE \(
        \phi \;\gets\; \phi \;-\; \eta \,\nabla_{\phi}\,\mathcal{L}(\phi, \phi^-)
      \)
      \STATE \(
        \phi^- \;\gets\; \mathrm{stopgrad}\Bigl(\,\mu\,\phi^- \;+\; \bigl(1 - \mu\bigr)\,\phi\Bigr)
      \)
    \UNTIL{convergence}
  \end{algorithmic}
\end{algorithm}
The core of this training algorithm is to teach the model \(f_\phi\) to produce the same output in one forward pass at a higher noise level as an exponential moving average (EMA) \qq{teacher} model \(f_{\phi^-}\) would produce at a slightly lower noise level. Each iteration begins by sampling a pair \((\mathbf{\theta},y)\) from the dataset $\mathcal{D}$ and drawing a noise index \(n\) uniformly from \(\{1,\dots,N\}\). A Gaussian vector \(z\) is scaled by \(t_{n+1}\) and \(t_n\) to create two noisy versions: \(\mathbf{\theta} + t_{n+1}z\) (more corrupted) and \(\mathbf{\theta} + t_n z\) (slightly less corrupted). The student network \(f_\phi\) is evaluated on the more corrupted input, while the teacher network is evaluated on the slightly cleaner input. The distance 
\[
d\bigl(f_\phi(\mathbf{\theta} + t_{n+1}z,\,t_{n+1},\,y),\;f_{\phi^-}(\mathbf{\theta} + t_n z,\,t_n,\,y)\bigr)
\]
between these two outputs becomes the consistency loss \(\mathcal{L}(\phi,\phi^-)\). A gradient descent step on \(\phi\) minimizes this loss so that \(f_\phi\) learns to match the teacher’s output across noise levels, and then the teacher’s weights are updated to remain a smoothed version of the student. 
By repeating this process over random noise levels, \(f_\phi\) learns a single‐shot mapping from any noisy sample \(\bigl(\mathbf{\theta} + t z,\,t,\,y\bigr)\) to a clean output that closely approximates a full, multi‐step diffusion trajectory, thus dramatically reducing sampling time while preserving conditional fidelity.

\begin{figure*}[t!]
    \centering
    \includegraphics[width=0.9\textwidth]{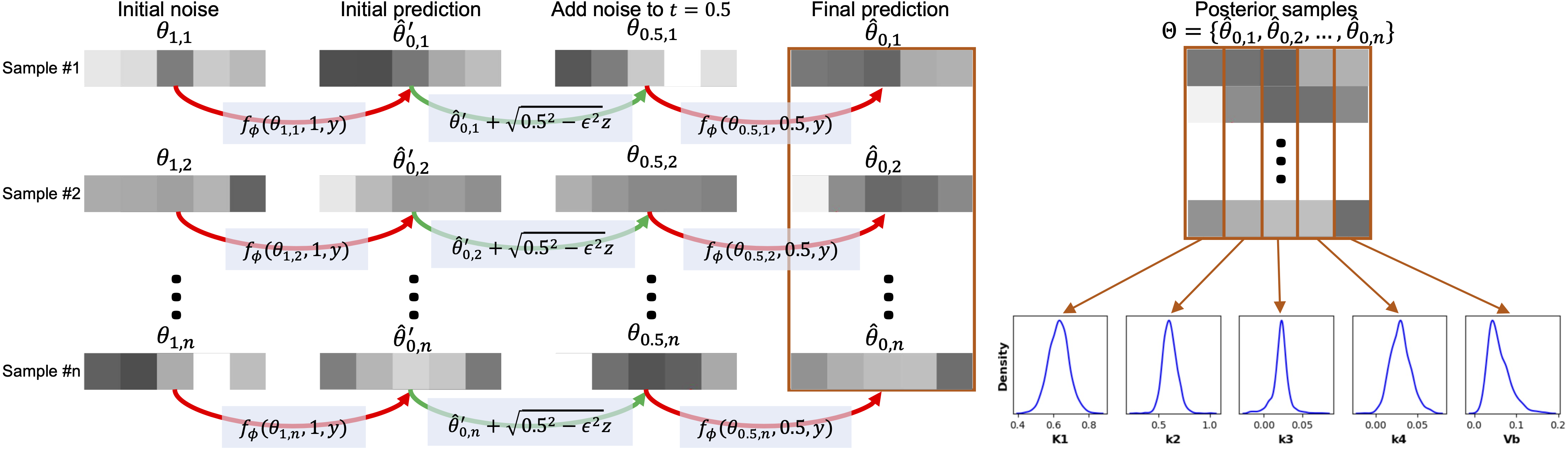}
    \caption{\textbf{Multistep consistency sampling for posterior kinetic parameter estimation.} 
    To draw a posterior sample, a zero-mean Gaussian vector shaped like the kinetic-parameter vector (e.g., $\mathbf{\theta}_{1,1}$) is generated and together with the measured TAC and AIF $y$ is passed through the consistency model $f_{\phi}(\cdot)$ at the highest noise level ($t = 1$), producing a coarse denoised estimate $\hat{\mathbf{\theta}}'_{0,1}$. This estimate is corrupted to the intermediate noise level ($t = 0.5$) to form $\mathbf{\theta}_{0.5,1}$, which is then fed back into $f_{\phi}(\cdot)$ to obtain the refined sample $\hat{\mathbf{\theta}}_{0,1}$. Repeating this procedure $n$ times with independent noise draws yields a posterior sample, treated as independent draws from the posterior distribution $p(\mathbf{\theta}|y)$.
    }
    \label{fig: Multistep consistency sampling}
\end{figure*}

Sampling with a trained consistency model proceeds by transforming a Gaussian noise sample to a posterior sample based on the condition \(y\). The sampling algorithm follows Multistep Consistency Sampling in the original paper \cite{song2023consistency} but with adaptations (See \textbf{Algorithm} \ref{alg:multistep-consistency-sampling}).
Fig. \ref{fig: Multistep consistency sampling} illustrates Multistep Consistency Sampling.
The procedure is as follows. First, draw a random Gaussian noise vector \(\mathbf{\theta}_{T}\) at the largest noise scale \(T\). Apply the consistency model \(f_{\phi}\) once to this noisy input \(\bigl(\mathbf{\theta}_{T},\,T,\,y\bigr)\). Since the network has been trained to “jump” from any noise level directly toward a clean sample conditioning on \(y\), this first evaluation yields an approximately denoised estimate \(\hat{\mathbf{\theta}}_{0}\) based on the conditioning information.
Next, iterate over a strictly decreasing sequence of noise levels
\[
t_{1} > t_{2} > \cdots > t_{N-1}.
\]
At each step \(n\), inject a fresh Gaussian noise vector \(z \sim \mathcal{N}(0,I)\) into the current estimate \(\hat{\mathbf{\theta}}_{0}\) by computing
\[
\mathbf{\theta}_{t_{n}} \;=\; \hat{\mathbf{\theta}}_{0} \;+\; t_{n}z,
\]
which simulates corruption of \(\hat{\mathbf{\theta}}_{0}\) to noise level \(t_{n}\). Then feed \(\bigl(\mathbf{\theta}_{t_{n}},\,t_{n},\,y\bigr)\) through \(f_{\phi}\) to refine the denoising in a way that remains consistent with the earlier, coarser estimate and the condition \(y\). In other words,
\[
\hat{\mathbf{\theta}}_{0} \;\gets\; f_{\phi}\bigl(\mathbf{\theta}_{t_{n}},\,t_{n},\,y\bigr).
\]
After processing the final noise level $N-1$, the last output \(\hat{\mathbf{\theta}}_{0}\) is taken as the final, posterior sample \(\mathbf{\theta}|y\).

\begin{algorithm}[htp]
  \caption{Multistep Consistency Sampling}
  \label{alg:multistep-consistency-sampling}
  \begin{algorithmic}[0]
    \REQUIRE Conditional consistency model $f_{\phi}(\,\cdot\,,\,\cdot\,,\,\cdot\,)$, sequence of noise levels $t_1 > t_{2} > \cdots > t_{N-1}$, initial noisy sample $\mathbf{\theta}_{T}$, auxiliary information $y$
    \STATE $\hat{\mathbf{\theta}}_0 \;\gets\; f_{\phi}\bigl(\mathbf{\theta}_{T},\,T, y)$
    \FOR{$n = 1$ \TO $N-1$}
      \STATE Sample $z \sim \mathcal{N}(0,\,I)$
      \STATE $\displaystyle
        \mathbf{\theta}_{t_{n}} \;\gets\; \hat{\mathbf{\theta}}_0 \;+\; t_{n}z
      $
      \STATE $\hat{\mathbf{\theta}}_0 \;\gets\; f_{\phi}\bigl(\mathbf{\theta}_{t_{n}},\,t_{n}, y)$
    \ENDFOR
    \STATE \textbf{Output:} $\hat{\theta}_0$
  \end{algorithmic}
\end{algorithm}

\subsubsection{Model Architectures}
\begin{figure*}[t!]
    \centering
    \includegraphics[width=0.72\textwidth]{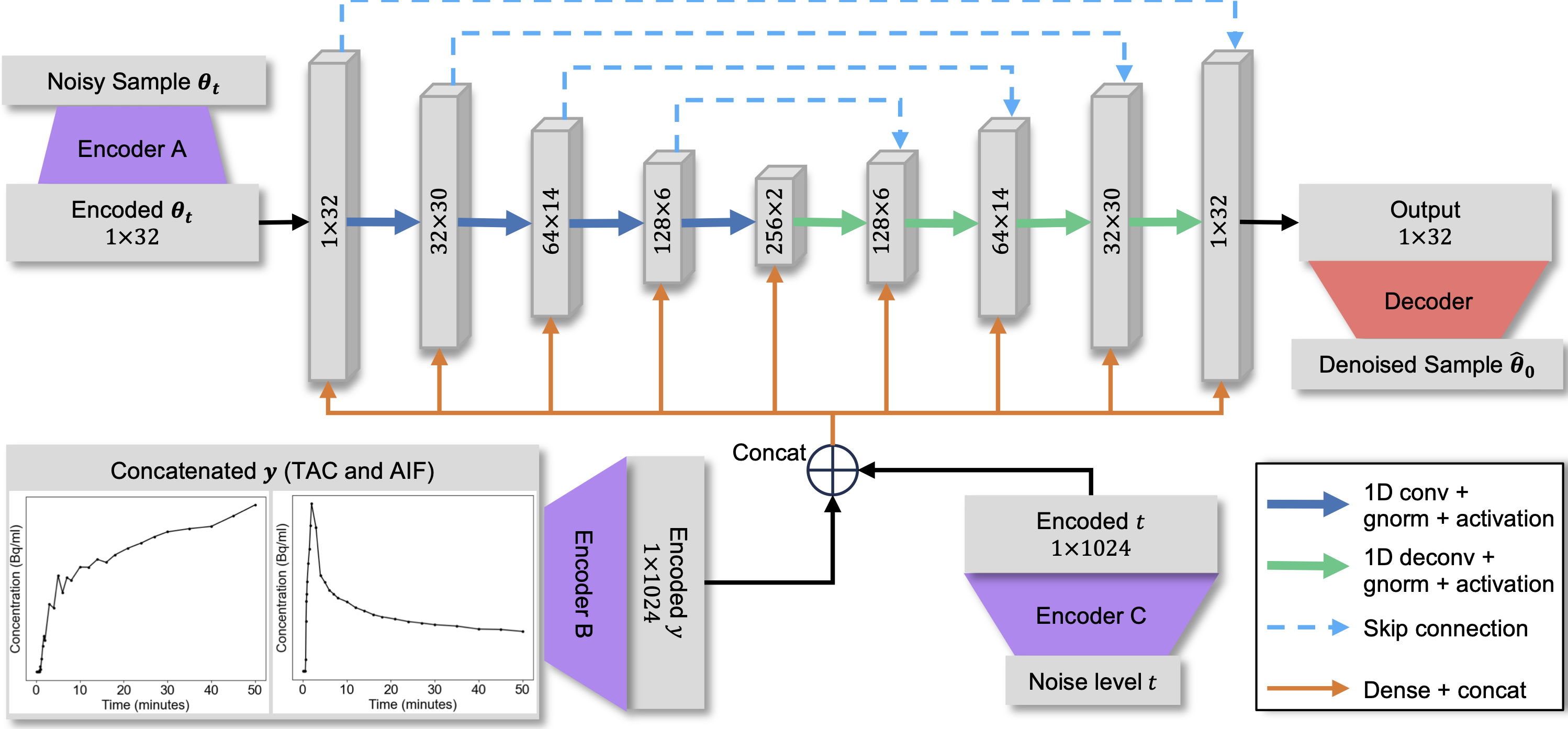}
    \caption{\textbf{Architecture of the 1D U-Net used in the consistency model for kinetic‐parameter posterior estimation.} The network combines three encoders: encoder A and B are dense layer-based encoders for the noisy kinetic parameter values $\mathbf{\theta}_t$ and concatenated dynamic data $y$, respectively, and the noise level $t$ is encoded by applying a Gaussian Fourier projection. Encoded $y$ and $t$ are concatenated and injected at each downsampling (blue arrow) and upsampling stage (green arrow). Skip connections (dashed blue) link corresponding encoder and decoder blocks. We use this neural network as the $f_\phi(\cdot)$.
    }
    \label{fig: Unet}
\end{figure*}

The building block of the proposed conditional consistency model is a 1D U‐Net (Fig. \ref{fig: Unet}) that processes two inputs: the noisy kinetic parameter vector \(\mathbf{\theta}_t \in \mathbb{R}^{\mathbf{\theta}_{1\times n}}\) and the concatenated dynamic measurements \(y \in \mathbb{R}^{y_{1\times m}}\) consisting of a TAC and the corresponding AIF, together with a scalar noise level \(t\). First, \(\mathbf{\theta}_t\) is transformed to a vector of size 32 (a dimension determined empirically to balance representational capacity with computational efficiency) for the following 1D convolutions, while \(y\) is embedded through a dense layer. Simultaneously, \(t\) is embedded via Gaussian Fourier features followed by a linear projection. The two resulting embeddings are combined via element-wise SiLU activation \cite{elfwing2018sigmoid}. The SiLU is smooth and non-monotonic, with a non-zero derivative even for negative inputs, which often leads to faster convergence and improved gradient flow compared to ReLU or other piecewise‐linear activations. This combined embedding vector is then injected through a dense layer at every level of the U‐Net to condition all convolutional blocks on both \(y\) and \(t\). 
The encoder (downsampling) path consists of four successive 1D convolutional blocks with kernel size 3. The first block maps \(1 \to 32\) channels with stride 1, the second maps \(32 \to 64\) channels with stride 2, the third maps \(64 \to 128\) channels with stride 2, and the fourth maps \(128 \to 256\) channels with stride 2. After each convolution, the network adds the dense projection of the combined embedding, applies group normalization, and then applies the SiLU activation. Residual (skip) connections carry the output of each encoder block forward into the matching decoder stage. The decoder (upsampling) path mirrors this structure using transposed 1D convolutions: \(256 \to 128\), \(128 \to 64\), \(64 \to 32\), and finally \(32 \to 1\) channels. At each upsampling stage, the decoder concatenates the corresponding encoder feature map, injects the same combined embedding via a dense layer, applies group normalization, and uses SiLU activation to restore spatial resolution. A final dense layer projects the length of 32 feature map back to the original dimension \(\mathbf{\theta}_{1\times n}\), yielding a denoised $\hat{\mathbf{\theta}}_0$. 

\subsubsection{Parametrisation for Consistency Models}
To enforce the boundary condition of the CM (i.e., that the model’s output matches \(\mathbf{\theta}_0\) at \(t=0\)), we compute two noise level‐dependent scalar coefficients
\[
c_{\mathrm{skip}}(t) \;=\; \frac{\sigma_{\mathrm{data}}^2}{\,t^2 + \sigma_{\mathrm{data}}^2\,}
\qquad\text{,}\qquad
c_{\mathrm{out}}(t) \;=\; \frac{\sigma_{\mathrm{data}}\,t}{\sqrt{\,t^2 + \sigma_{\mathrm{data}}^2\,}},
\]
where $\sigma_{\mathrm{data}}$ represents the standard deviation of the clean data distribution for the target $\mathbf{\theta}$. The final output of the neural network is given by
\[
f_\phi(\mathbf{\theta}_t,t,y) \;=\; c_{\mathrm{skip}}(t)\,\mathbf{\theta}_0 \;+\; c_{\mathrm{out}}(t)\,F(\mathbf{\theta}_t,t,y),
\]
where $F(x_t,t,y)$ represents the 1D U-Net described in the last subsection. This formulation guarantees that at $t=0$, $c_{skip}(0)=1$ and $c_{out}(0)=0$, so $f_\phi(\mathbf{\theta}_t,t,y)=\mathbf{\theta}_0$. As $t$ increases, more emphasis is placed on $F(\mathbf{\theta}_t,t,y)$, allowing the output to rely on its learned denoised $\hat{\mathbf{\theta}}_0$.

\subsection{Application of  CM to Dynamic PET}

\subsubsection{Two-Tissue Compartment Model}\label{2TC}
The two-tissue compartment model describes the tracer kinetics of [\textsuperscript{18}F]FDG within a tissue of interest, partitioning the tracer into a free compartment, $C_f (t)$, and a compartment representing phosphorylated (i.e. metabolised) FDG in the form of FDG-6-phosphate, $C_m (t)$. Let $C_b (t)$ denote the AIF of the tracer \cite{phelps1979tomographic}. The net uptake of tracer from plasma into tissue compartment $C_f (t)$ is captured by the rate constant $K_1$, while $k_2$ represents efflux from this compartment back to plasma. The transfer of tracer from compartment $C_f (t)$ to $C_m (t)$ is governed by the rate constant $k_3$, representing phosphorylation of FDG, while the reverse phosphorylation (commonly assumed negligible) is represented by rate constant $k_4$. These processes are summarised by the following system of ordinary differential equations,
\begin{align*}
\frac{dC_f(t)}{dt} &= K_1\,C_b(t)\;-\;(k_2 + k_3)\,C_f(t)\;+\;k_4\,C_m(t), \\
\frac{dC_m(t)}{dt} &= k_3\,C_f(t)\;-\;k_4\,C_m(t).
\end{align*}

Although $C_f (t)$ and $C_m (t)$ describe tracer concentrations within tissue, the dynamic PET time series, $y(t)$, typically contains both tissue and a small intravascular contribution within the sampled image voxel(s). To account for the fraction of the voxel occupied by blood, we introduce a blood volume term, $V_b$, where $0\leq V_b\leq1$. The simulated PET signal is thus,
\[
C_t(t) = (1-V_b)(C_f(t)+C_m(t))+V_bC_{wb}(t).
\]
where $C_{wb}(t)$ represents the whole-blood radioactivity concentration. In this study, both $C_b(t)$ and $C_{wb}(t)$ were approximated using the image-derived input function extracted from the ascending aorta.
To account for measurement noise, a Gaussian noise model is introduced \cite{wu2002noise}, and the measured PET time series is assumed to be,
\[
    y(t) = C_t(t) + \ell\,\sigma_t\,\mathcal{N}(0,1),
\]
where $\sigma_t=\sqrt{\frac{C_t(t)\mathrm{exp}(-\lambda t)}{\Delta t}}\mathrm{exp}(\lambda t)$. $\Delta t$ is the duration of a given frame in the PET time series, $\lambda=\mathrm{ln}2/T_{1/2}$ is the decay constant for $\prescript{18}{}{\mathrm{F}}$, with $T_{1/2}=109.8$ min and $l$ is a constant that scales the noise level.

\subsubsection{Training Datasets}
The kinetic parameter vectors \((K_{1},k_{2},k_{3},k_{4},V_{b})\) were sampled uniformly from ranges reported in the PET literature for FDG in a wide variety of organs and tissues \cite{liu2021kinetic,sari2022first}. Specifically, we chose 
\[
\begin{aligned}
K_{1} &\in [0.001,\,1.00]\ \mathrm{mL\,min^{-1}\,cm^{-3}}, \\
k_{2} &\in [0.001,\,2.00]\ \mathrm{min^{-1}},
k_{3} \in [0.001,\,0.50]\ \mathrm{min^{-1}}, \\
k_{4} &\in [-0.10,\,0.10]\ \mathrm{min^{-1}},
V_{b} \in [0.03,\,0.20],
\end{aligned}
\]
where these intervals were found to encapsulate the vast majority of physiological tissues. 
We allowed $k_4$ to take negative values during network training. This allows for a robust estimation of the parameter distribution when the true value lies near the boundary of zero.
In other words, exposing the model to negative $k_4$ values ensures that network predictions around $k_4=0$ are within the training range and are reliable \cite{xu2020neural}, enabling the robust model selection analysis described in Section \ref{section: model selection}.

The AIF \(C_{b}(t)\) was simulated using the Feng model \cite{feng1993models},
\[
\begin{aligned}
C_{b}(t)=&
\bigl(\beta_{1}(t - t_{0})-\beta_{2}-\beta_{3}\bigr)\,e^{-\kappa_{1}(t - t_{0})}\\
&+\beta_{2}\,e^{-\kappa_{2}(t - t_{0})}
+\beta_{3}\,e^{-\kappa_{3}(t - t_{0})}, 
\end{aligned}
\]
where the seven parameters \((\beta_{1},\beta_{2},\beta_{3},\kappa_{1},\kappa_{2},\kappa_{3},t_{0})\) were drawn uniformly from ranges covering the full variability observed in eight oncological subjects
aged 43-82 years (Mean$\pm$SD: $70.53 \pm 10.36$), weighing 55-97 Kg (Mean$\pm$SD: $71.23 \pm 12.80$ Kg) acquired on the same Siemens Biograph Vision Quadra. All subjects received [\textsuperscript{18}F]FDG injection (Mean$\pm$SD: $146.67 \pm 11.50$ MBq) and were reconstructed using the same OSEM protocol described in Section \ref{section: real data}.
\[
\begin{aligned}
&\beta_{1}\in[4\times10^{4},\,4.7\times10^{5}],\quad
\beta_{2}\in[1\times10^{4},\,2\times10^{5}],\quad\\
&\beta_{3}\in[8\times10^{3},\,3\times10^{4}],\quad
\kappa_{1}\in[1,\,1\times10^{3}],\quad\\
&\kappa_{2}\in[0.03,\,3.60],\quad
\kappa_{3}\in[0.01,\,0.035],\quad
t_{0}\in[0,\,1.25]\ \mathrm{min}.
\end{aligned}
\]
Note that by explicitly varying $t_0$ during training, the model learns features invariant to tracer arrival delays, effectively performing an implicit delay correction during inference.
This parameterization ensured that the training set included diverse input function shapes (representing variations in injection speed/dispersion) and delays, enabling the model to learn from and generalize across realistic physiological conditions.
Noise was added to the noise-free simulation $C_t(t)$ via the noise model described in Section \ref{2TC} with noise level $l$ randomly drawn from $3$ to $7$. The range of noise levels was also derived from the clinical dynamic FDG PET data from the eight subjects. 

The noisy measured value is then
\[
\tilde{y}(t_{i})=\max\{\,0,\;y(t_{i})\},
\]
with negative values clipped to zero. For each draw of the 13 parameters, we solved the two‐tissue ODE system over 35 time points spanning \(5\)s to \(3000\)s (0.083–50min) to produce simulated noiseless curves, \(C_t(t)\) and \(C_{b}(t)\), then applied the noise model to obtain \(\tilde{y}(t)\). The final dataset consists of 500,000 training samples and 40,000 validation samples (each sample comprises a 35‐point noisy TAC \(\tilde{y}\) concatenated with the corresponding 35‐point \(C_{b}\) as input, paired with the true 5‐element kinetic parameter vector). This design ensures that the synthetic data span the full range of physiological values and noise variability observed in human FDG-PET kinetics and arterial inputs.

\subsubsection{Training Details}\label{subsec: training details}
The 1D U-Net was trained using the AdamW optimizer with a batch size of 200,000. Training was scheduled for a maximum of 80,000 epochs. The validation loss was observed to stabilize and plateau after 150,000 iterations. Consequently, the final model weights were selected from epoch 60,200 (approx. 150,500 iterations), as this checkpoint corresponded to the minimum validation error.
A separate teacher model was trained via an EMA of the student parameters with a fixed decay rate of 0.99. At each iteration, two adjacent noise levels drawn uniformly from a fixed schedule of $N_k=50$ steps, were used to compute the consistency loss between the student and teacher outputs, and the teacher parameters were updated.
For comparison, we tested an adaptive schedule in which $N_k$ and EMA $\mu$ started at small values and were gradually increased to the preset maximum $N_k$ and $\mu$ over the course of training, allowing coarser noise discretization early on and finer resolution later. We also tested various choices for $N_k$ (from 2 to 200). Finally, we found that the fixed 50-step noise schedule and fixed EMA $\mu$ yielded the best validation performance. 
We trained and evaluated all deep learning models on a single node equipped with 4 NVIDIA A100 GPUs (80GB VRAM) using data parallelisation.

\subsection{Performance Evaluation}
\subsubsection{Reference Method: MCMC}
Posterior ground truth was generated with the adaptive Hamiltonian Monte Carlo algorithm implemented in \textit{PyMC} \cite{hoffman2014no,abril2023pymc}.
For inference we ran four independent chains, each with 10,000 warm-up iterations, followed by 200,000 posterior draws per chain. A target acceptance probability of $0.9$ was retained, and the sampler’s automatic jittering of initial values ensured robust exploration even for poorly scaled parameters. To ensure the MCMC samples served as a reliable ground truth for the posterior distribution, strict convergence criteria were enforced. Convergence was confirmed only when all kinetic parameters satisfied the Gelman-Rubin convergence statistic $\hat{R} < 1.01$ \cite{vehtari2021rank} and the effective sample size exceeded 40,000. Chains failing these strict diagnostics were discarded and re-sampled with increased warm-up phases to guarantee that the MCMC reference represented the true stationary distribution.

\subsubsection{Compared Methods}
We benchmarked our CM against ABC and four deep learning baselines: CVAE, DDPM, score-based generative modeling through stochastic differential equations (SBD), and conditional Wasserstein generative adversarial network with gradient penalty (GAN).

ABC provides a likelihood‑free reference rooted in forward simulation: kinetic parameter vectors are drawn from the prior, TACs are simulated with the compartmental model, and a draw is retained only when the simulated TAC lies within a preset distance from the measured one \cite{sunnaaker2013approximate}. The accepted draws form an empirical posterior but, because proposals are not guided toward high‑probability regions, the acceptance rate decays rapidly with dimensionality. In our study, we instantiate ABC for dynamic PET using PET-ABC \cite{fan2021pet}.
The DDPM benchmarks posterior estimation by learning the reverse of a 1000‑step Gaussian noise process that maps a standard normal prior to the conditional posterior $p(\theta|y)$. During training, the network reconstructs the added noise at each diffusion level, optimizing a mean squared error that has a closed‑form connection to variational inference \cite{ho2020denoising}. We implemented the denoiser with the same 1‑D U‑Net backbone, depth and parameter count as CM to ensure capacity parity. Inference follows the original reverse chain to generate samples, incurring 1000 evaluations per draw.
SBD uses a continuous stochastic differential equation (SDE) whose forward form gradually perturbs data toward Gaussian noise, while a learned score function drives the reverse SDE back to the posterior \cite{song2020score}. Training minimizes a denoising score matching loss that directly estimates the score $\nabla_{\theta_t}\log p_t(\theta_t\mid y)$ at different noise levels. A U‑Net identical to the one used in our CM was used. Sampling integrates the reverse SDE with an Euler–Maruyama solver for 1000 time steps. 
For CVAE, two fully connected encoders map $(\theta,y)$ and $y$ alone to the mean and log‑variance of a latent Gaussian. Reparameterised noise is then concatenated with $y$ and passed through a fully connected decoder to reconstruct a kinetic parameter sample. For more details about CVAE, readers are directed to \cite{liu2023posterior}. We reused the CVAE-dual-decoder described in this reference. The three networks (i.e., two encoders and one decoder) were up‑sized so the number of learnable parameters matches the number of parameters in the U-Net used in CM.
The GAN trains a generator $G_{\mathbf{\theta}}$ to transform a Gaussian noise vector concatenated with the 70‑element TAC–AIF vector into plausible parameter samples, and a critic $D_{\psi}$ that enforces the Wasserstein distance with a gradient‑penalty regulariser. After convergence, posterior draws require only a single forward pass through $G_{\mathbf{\theta}}$. Readers are directed to \cite{gulrajani2017improved} for more details. The generator and critic adopt encoder–decoder U‑Nets whose width gives the total number of parameters comparable to CM and the diffusion baselines. 

\subsubsection{Performance Metrics}
First, the dissimilarity between the posterior produced by method $m$, $q_m(\theta|y)$, and the reference MCMC posterior, $p_{MCMC}(\theta|y)$, was measured by an empirical Kullback–Leibler divergence,
\[
\widehat{D}_{\mathrm{KL}}\!\bigl(p_{\text{MCMC}}\;\|\;q_m\bigr)
  = \frac{1}{N}\sum_{i=1}^{N}
    \log\frac{\hat{p}_{\text{MCMC}}(\theta_i)}
              {\hat{q}_m(\theta_i)},
\qquad \theta_i\sim p_{\text{MCMC}},
\]
where \(\hat{p}_{\text{MCMC}}\) and \(\hat{q}_m\) are kernel density estimates built from posterior samples, and $N$ is the sample size.
Second, the accuracy of the posterior mean was summarized by the absolute percentage error (APE) with respect to the ground-truth kinetic parameter $\theta^*$,
\[
\mathrm{APE}=\,
\frac{\lvert \mu_m-\theta^{\ast}\rvert}{\theta^{\ast}},
\qquad
\mu_m=\frac{1}{N}\sum_{i=1}^{N}\theta_i .
\]
Third, the quality of uncertainty estimates was assessed by the relative standard error (RSE) in posterior standard error with respect to that of MCMC,
\[
\mathrm{RSE}=\,
\frac{\lvert \sigma_m-\sigma_{\text{MCMC}}\rvert}{\sigma_{\text{MCMC}}},
\qquad
\sigma_m^{2}=\frac{1}{N}\sum_{i=1}^{N}(\theta_i-\mu_m)^{2},
\]
where \(\sigma_{\text{MCMC}}\) is the standard error of the MCMC posterior.  

\subsection{Dynamic [\textsuperscript{18}F]FDG Total-Body Study}\label{section: real data}
A 54 year old male subject with a pancreatic neuroendocrine tumour underwent a 50-min dynamic total-body PET scan on a Siemens Biograph Vision Quadra (Siemens Healthineers, Knoxville, USA) commencing with intravenous administration of 150 MBq of [\textsuperscript{18}F]FDG. This scan was part of a larger study approved by the Northern Sydney Local Health District Human Research Ethics Committee. Prior written informed consent to participate in this study was obtained from the subject.

List-mode data were histogrammed into 35 dynamic frames, each with matrix size $220\times220\times645$. Normalization, random, attenuation, decay and scatter corrections were applied and the data were reconstructed using ordinary Poisson ordered subsets expectation maximization algorithm (OP-OSEM, 8 iterations, 5 subsets), with time-of-flight weighting and a maximum ring difference of 322. 
Spatiotemporal HYPR-LR filtering \cite{christian2010dynamic} was applied to the dynamic series.
The AIF was extracted from the ascending aorta. First, a three-dimensional mask of the aorta was produced on the native CT images using MOOSE segmentation \cite{sundar2022fully}, and the mask was then morphologically eroded (two-voxel radius) to minimize partial volume contamination. The whole-blood AIF was taken as the mean activity concentration within this eroded region for each frame.
Previous validation studies have demonstrated that ascending aorta IDIFs obtained with erosion strategies provide robust kinetic estimates effectively equivalent to arterial sampling \cite{de2006comparison}.
Using the AIF and reconstructed voxel-based TACs as input, posteriors of the micro-parameters, $K_1, k_2, k_3$ and $k_4$ were estimated using CM. Then, the net influx rate, $K_i$, was computed  for every voxel as $K_i=K_1k_3/(k_2+k_3)$ to generate parametric images of $K_i$ which were compared with voxel-based Patlak $K_i$ images.

\section{Results}
\begin{figure*}[btp]
    \centering
    \includegraphics[width=0.9\textwidth]{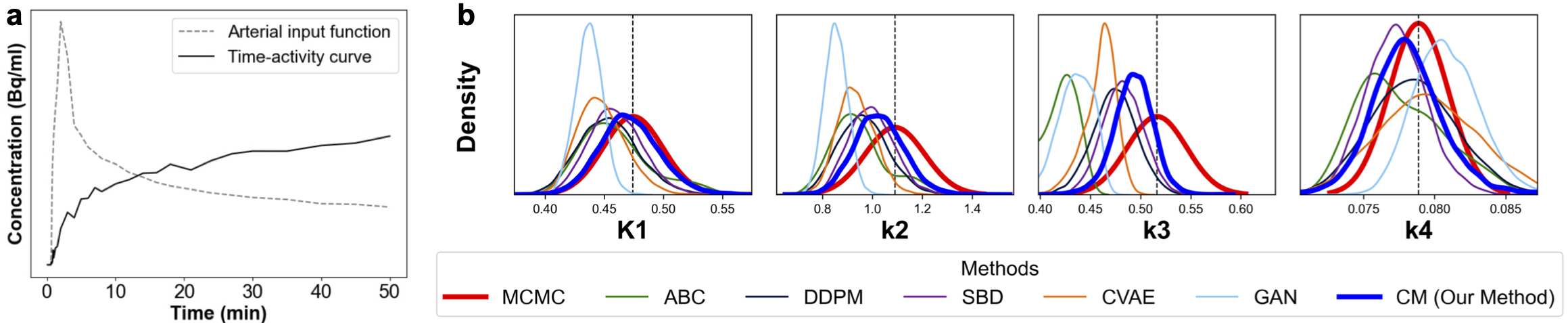}
    \caption{\textbf{Example posterior inference for a single voxel.} (a) A simulated time–activity curve is shown with the corresponding AIF. (b) The resulting posterior distributions of the kinetic parameters, predicted by the proposed consistency model (blue), are contrasted with those obtained from deep learning baselines, approximate Bayesian computation, and the reference Markov-chain Monte Carlo (MCMC) method (red). The black dashed lines represent the truth value.
    }
    \label{fig: Posterior estimation example}
\end{figure*}

\subsection{Evaluation on Simulation Data}
We first evaluated our proposed CM on simulated testing data. Fig. \ref{fig: Posterior estimation example} shows an example where the posteriors were estimated on one TAC-AIF pair from the testing dataset.
The CM yields posteriors that closely approximate the ones predicted by the MCMC, particularly for $K_1$. For the other parameters ($k_2, k_3, k_4$), while minor distributional shifts or variances are observed compared to the MCMC prediction, the CM posterior consistently encapsulates the ground truth values (dashed lines) within its credible interval (CrI).
The ABC posterior retains similar means but displays noticeably larger tails, indicating larger variance.
Among the diffusion based methods, the DDPM produces slightly wider peaks, particularly for $K_1$, $k_2$, and $k_4$, whereas the SBD exhibits narrower shoulders, suggesting slight overdispersion and underdispersion, respectively.
The CVAE compresses each distribution, implying a tendency to underestimate variance for $K_1$, $k_2$, and $k_4$, while the GAN shifts the means of all kinetic parameters and narrows their peaks, suggesting both bias and decreased uncertainty. 

\begin{figure}[htp]
    \centering
    \includegraphics[width=\columnwidth]{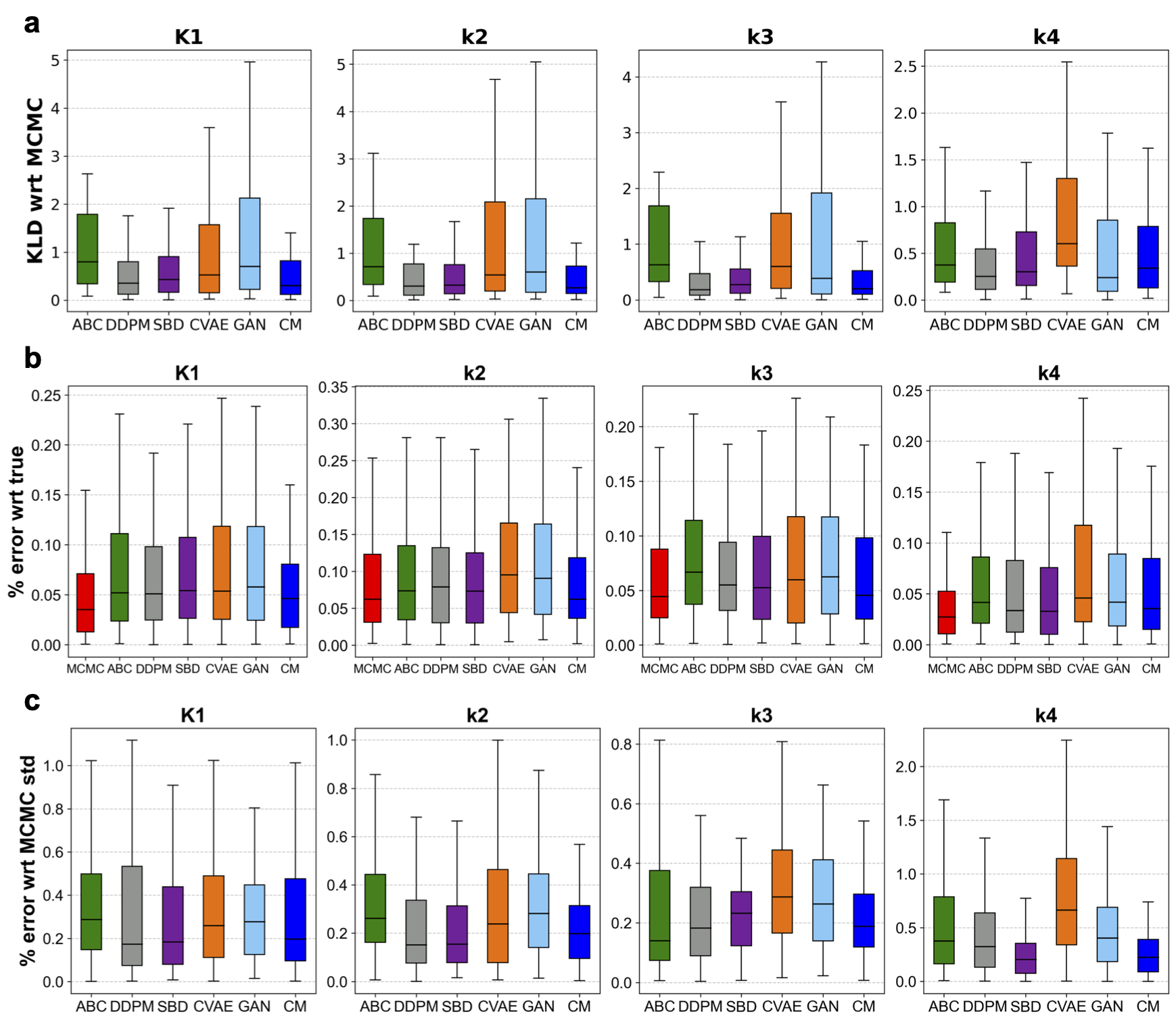}
    \caption{\textbf{Quantitative evaluation on 100 simulated TACs.} For each trial, a kinetic parameter vector was drawn from the pre-assumed physiological range, a noisy TAC was generated with a two-tissue compartment model, and posteriors were estimated by the proposed CM, the reference MCMC, and five baselines—ABC, DDPM, SBD, CVAE, and GAN. (a) Kullback–Leibler divergence between each method’s posterior and the MCMC posterior. (b) Absolute percentage error between the ground-truth parameter and the posterior mean. (c) Percentage error of the posterior standard deviation relative to that obtained with MCMC.
    }
    \label{fig: Evaluation on simulation}
\end{figure}

Fig. \ref{fig: Evaluation on simulation} extends the analysis of Fig. \ref{fig: Posterior estimation example} to 100 independent TAC–AIF pairs. For every pair we compared the posteriors produced by our CM with those from MCMC and five baselines—ABC, DDPM, SBD, CVAE and GAN. Panel (a) plots the Kullback–Leibler divergence (KLD) between each method’s posterior and the MCMC reference. The CM attained the lowest KLD median for $K_1$, the second lowest for $k_2$ and $k_3$, and yielded a KLD for $k_4$ comparable to ABC, outperforming both the CVAE and GAN.
Panel (b) focuses on point-estimate accuracy using the absolute percentage error (APE) between the posterior mean and the ground-truth parameters. We include MCMC in this comparison as a benchmark. Because each posterior is conditioned on one noisy TAC–AIF realization, even MCMC yields a non-zero APE—the posterior mean need not coincide exactly with the latent kinetic vector. Consequently, boxplots that cluster near the MCMC box denote methods whose bias is comparable to MCMC. 
In this setting, the error distribution of CM most closely approximates that of the reference MCMC method for $K_1$, $k_2$ and $k_3$, indicating minimal bias in the posterior mean estimates. However, for $k_4$, no approach achieves errors near the MCMC benchmark; instead, diffusion‐based methods (DDPM and SBD) outperform CM.
Panel (c) examines uncertainty estimation, reporting for each kinetic parameter the percentage error between a method’s posterior standard deviation and that obtained with MCMC. Accurate uncertainty estimation is critical, as downstream model selection depends on reliable CrIs. 
For $K_1$, the CM and the diffusion‐based methods (DDPM, SBD) exhibit median absolute errors of approximately $20\%$, whereas all other approaches show larger medians. 
A similar pattern holds for $k_2$: DDPM and SBD remain closest to MCMC, with CM displaying a slightly higher median error. In the case of $k_3$, ABC achieves the lowest median error, and DDPM and SBD follow within less than one tenth, while CVAE and GAN overshoot by roughly twice that margin. Uncertainty estimation proves most demanding for $k_4$. SBD yields the best alignment with MCMC, followed by CM. DDPM and ABC produce comparable median errors, and CVAE and GAN deviate the most.

\begin{figure}[htp]
    \centering
    \includegraphics[width=\columnwidth]{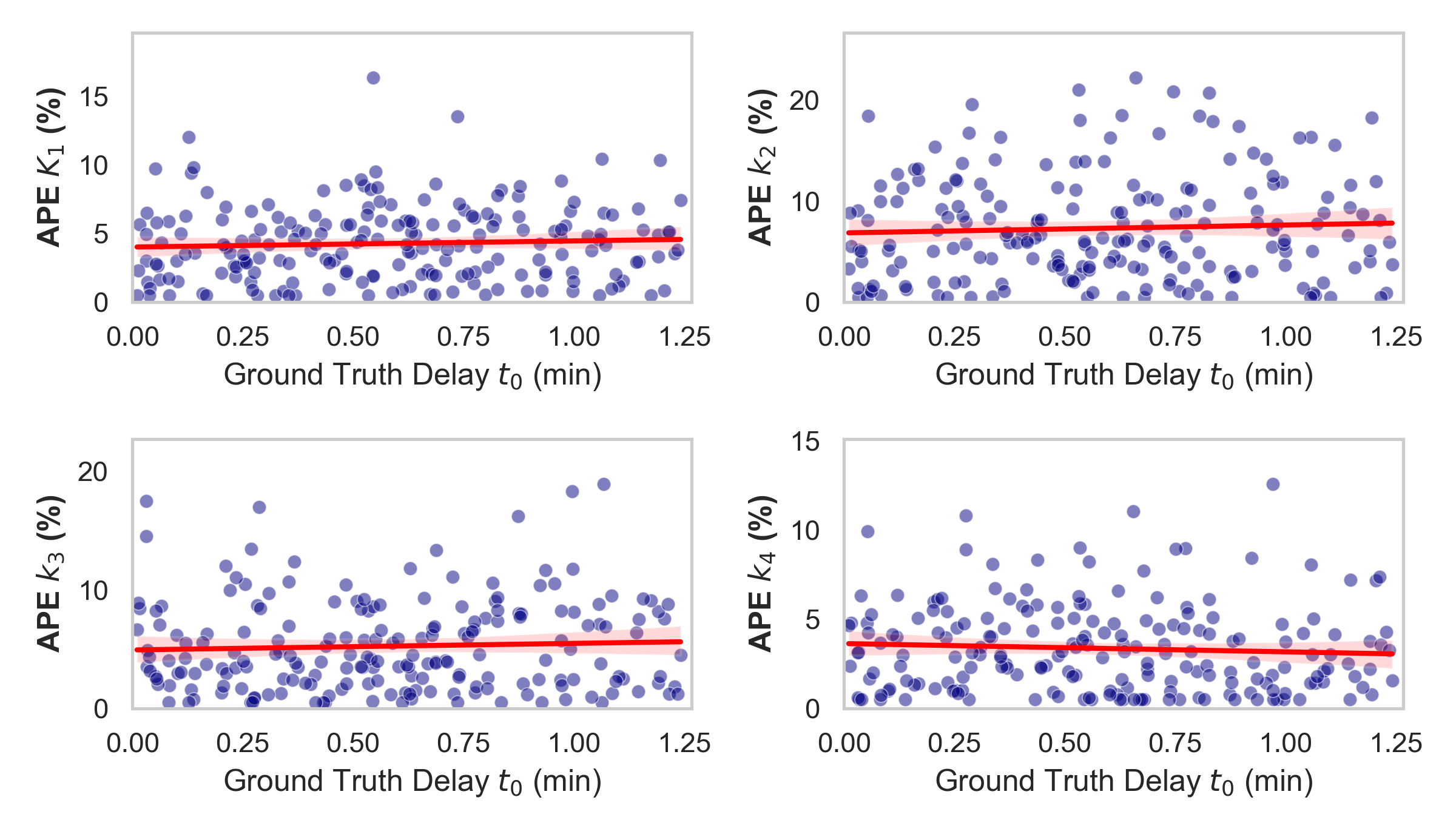}
    \caption{\textbf{Robustness of Kinetic Parameter Estimation to Tracer Arrival Delays.} Analysis of estimation accuracy with respect to tracer arrival delay ($t_0$). The absolute percentage error (APE) for each kinetic parameter is plotted against the ground-truth delay time for test samples. 
    }
    \label{fig: Params error vs tracer arrival time}
\end{figure}

To assess the robustness of the CM to variable tracer arrival times, we stratified the estimation error for kinetic parameters ($K_1, k_2, k_3, k_4$) against the ground-truth delay $t_0$ present in the test samples. As shown in Fig. \ref{fig: Params error vs tracer arrival time}, the APE remains stable across the delay range ($0$ to $1.25$ min) for all parameters, confirming that the CM successfully marginalizes over the delay parameter without requiring explicit temporal alignment of the AIF.

\subsection{Evaluation on Clinical Data}
\begin{figure}[htp]
    \centering
    \includegraphics[width=\columnwidth]{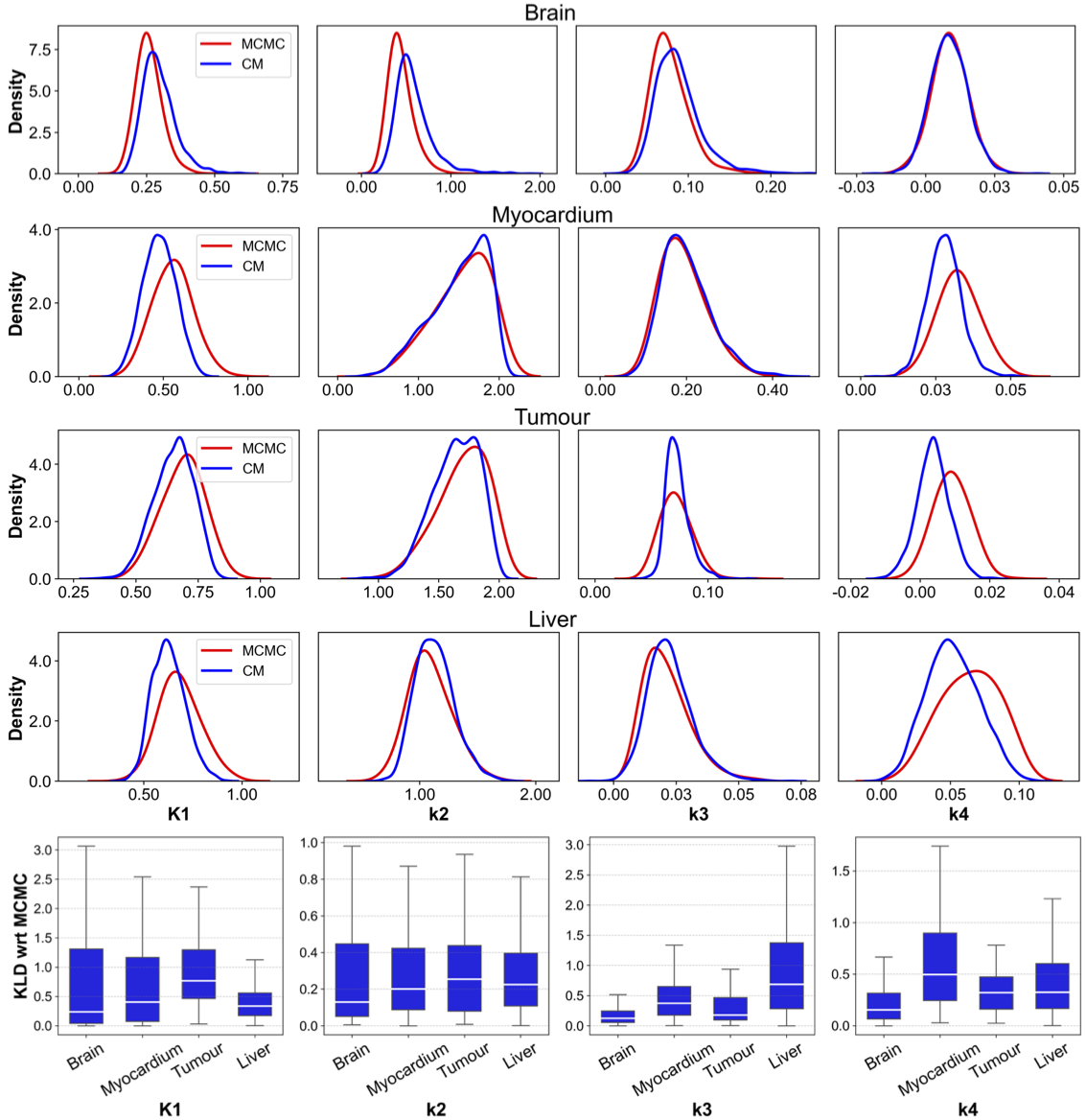}
    \caption{\textbf{Evaluation of voxel-wise posterior inference in clinical tissues.} Rows 1–4: Posterior distributions of kinetic parameters ($K_1, k_2, k_3, k_4$) for representative single voxels extracted from four regions: Brain (cerebral cortex), Myocardium, Tumour (hepatic lesion), and healthy Liver. Row 5: Boxplots of the KLD between posteriors predicted by CM and MCMC, calculated over 100 voxels within each region.
    }
    \label{fig: Evaluation on real data}
\end{figure}

The proposed CM was assessed on the clinical [$^{18}$F]FDG total-body PET scan described in Section \ref{section: real data}. To demonstrate clinical relevance across a range of organs, we evaluated posterior inference in four distinct targets: cerebral gray matter, myocardium, a hepatic tumour, and healthy liver.
As illustrated in Fig. \ref{fig: Evaluation on real data} rows 1–4, the CM posterior distributions (blue) exhibit agreement with the MCMC reference (red) across all kinetic parameters. 
In high-signal regions such as the cerebral cortex and myocardium, the model consistently resolves the sharp posterior modes associated with rapid tracer delivery and uptake.  
In the hepatic tumour, the CM accurately captures the irreversible trapping characteristic of malignant lesions, showing high probability density for $k_4=0$. For the healthy liver background, the CM correctly differentiates the distinct metabolic kinetics (lower $k_3$, higher $k_4$) from the adjacent tumour tissue.
To substantiate these observations quantitatively, Fig. \ref{fig: Evaluation on real data} row 5 presents the KLD between posteriors predicted by CM and MCMC, computed with 100 voxels for each region. The results indicate consistently low divergence (median KLD generally $< 0.5$) across distinct physiological environments, from the high-noise tumour to the high-uptake myocardium. This confirms that the CM's predictive accuracy is not an artifact of selected examples but reflects a robust generalized ability to recover the posterior distribution.

\subsection{Total-Body Parametric Imaging}
\begin{figure*}[t]
    \centering
    \includegraphics[width=0.77\textwidth]{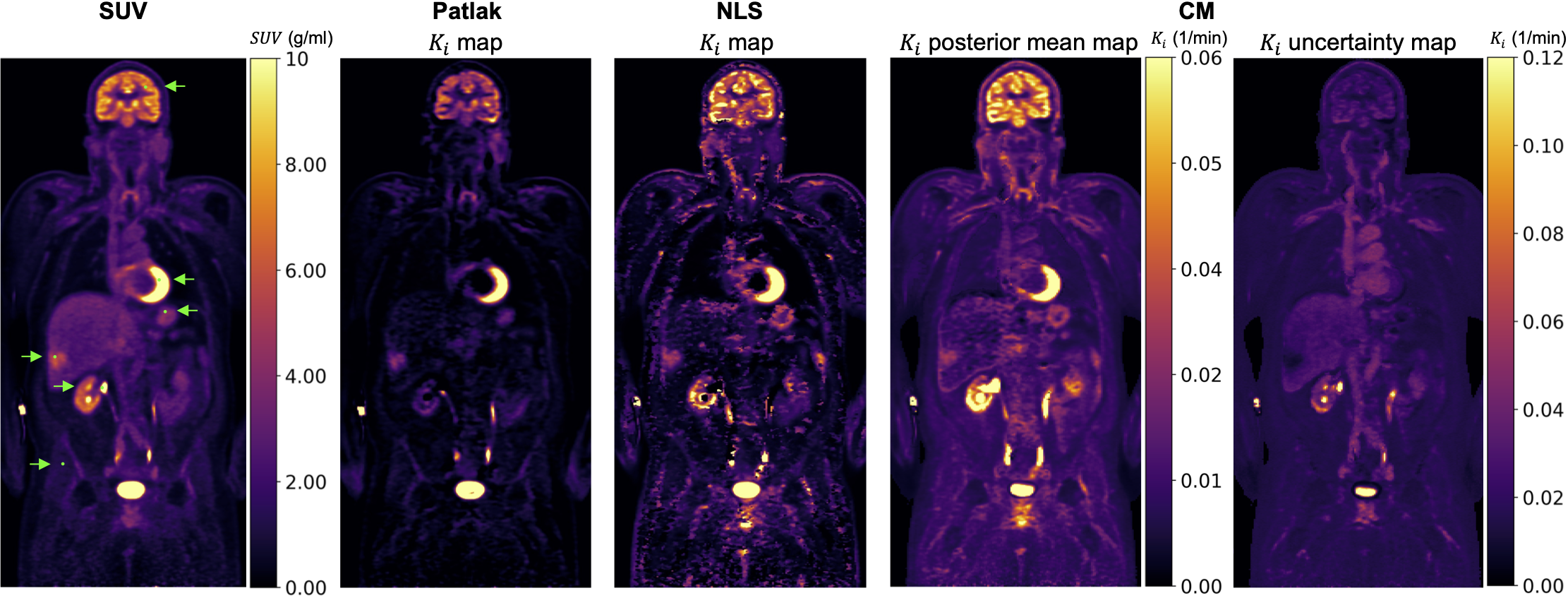}
    \caption{\textbf{SUV reference image and $K_i$ parametric maps from a dynamic TB-PET data.}
    From left to right: static clinical SUV image (average of the last frame, 45–50 min); conventional Patlak $K_i$ map (linear fit over $t^* > 18$ min, last 8 frames); $K_i$ map calculated by fitting the full two tissue compartment models using nonlinear least squares; $K_i$ map estimated by the proposed consistency model (CM); $K_i$ uncertainty map (i.e., width of 95\% CrI). 
    \label{fig:tb-parametric-imaging-treatment}}
\end{figure*}

Fig. \ref{fig:tb-parametric-imaging-treatment} presents the SUV and $K_i$ images calculated with Patlak analysis, nonlinear least squares (NLS), and the CM.  
The SUV image shows the expected FDG biodistribution: high uptake of FDG in cortical brain and myocardium, and renal cortices, and excretion into the urinary bladder. Intermediate uptake is observed throughout the liver and skeletal muscle; and very low uptake in the lungs. A focal area of high uptake is evident at the inferior right hepatic margin corresponding to a known metastatic tumour.
Qualitatively, the Patlak map preserves the expected physiological pattern of high $K_i$ in cortical grey matter of the brain, myocardium and renal cortex; intermediate uptake in liver and skeletal muscle, and low activity in lung parenchyma, but exhibits pronounced streak artefacts and amplified noise in low count regions.

The CM derived $K_i$ image generally agrees with the Patlak image.
High influx rates are clearly delineated in the same organs and tissues as the Patlak $K_i$ image.
The hyper‑metabolic tumour in the right lobe of the liver is evident but without the streak artefacts and amplified noise seen in the Patlak image. Across low‑count regions like pelvis and soft tissues, the CM image exhibits reduced variance, indicating a more favorable noise–bias trade‑off than Patlak.
To investigate the differences between CM and Patlak, we computed ROI-averaged TACs and inferred the kinetic parameters using the Patlak, CM and MCMC.
Patlak yields lower estimates compared to the MCMC reference. In the brain ($K_i^{\text{Patlak}}{=}0.0233$, $K_i^{\text{MCMC}}{=}0.0347$) and myocardium ($K_i^{\text{Patlak}}{=}0.0733$, $K_i^{\text{MCMC}}{=}0.0912$), the Patlak estimates are notably lower, whereas the CM ($0.0369$ and $0.0881$, respectively) is closer to the MCMC reference. This distinction is also critical in the liver where the Patlak predicts $K_i=0.0040$ while the CM ($0.0090$) is closer to the truth predicted by MCMC ($0.0078$). 
This highlights Patlak's limitations: assuming irreversible uptake ($k_4=0$) while neglecting fractional blood volume ($v_b$) and tracer arrival delay, thereby leading to biased estimates.
For comparison with a standard full-model approach, voxel-wise NLS estimation of the reversible two tissue compartment model was also performed. While NLS aligns generally with the CM estimates in high-count regions (e.g., brain, heart, kidneys), it exhibits higher variance in lower-count regions (e.g., liver, muscle) compared to the CM posterior means.
The accompanying $K_i$ uncertainty map depicts, for every voxel, the width of the 95\% CrI derived from the voxel-wise posterior distributions, elucidating the spatial distribution of estimation uncertainty.
The highest uncertainty is confined to regions exhibiting intense tracer activity and rapid kinetics, principally the renal cortices and urinary bladder where the 2-tissue compartment model does not apply.
In contrast, the cerebral cortex and skeletal muscles demonstrate low uncertainty, and the tumour in the liver likewise shows low levels of uncertainty, indicating a high degree of confidence in the parameter estimates for these regions.
Such explicit, spatially resolved confidence information is unavailable from SUV, NLS or Patlak analyses and constitutes a key advantage of the proposed CM framework.

\begin{figure}[htp]
    \centering
    \includegraphics[width=0.72\columnwidth]{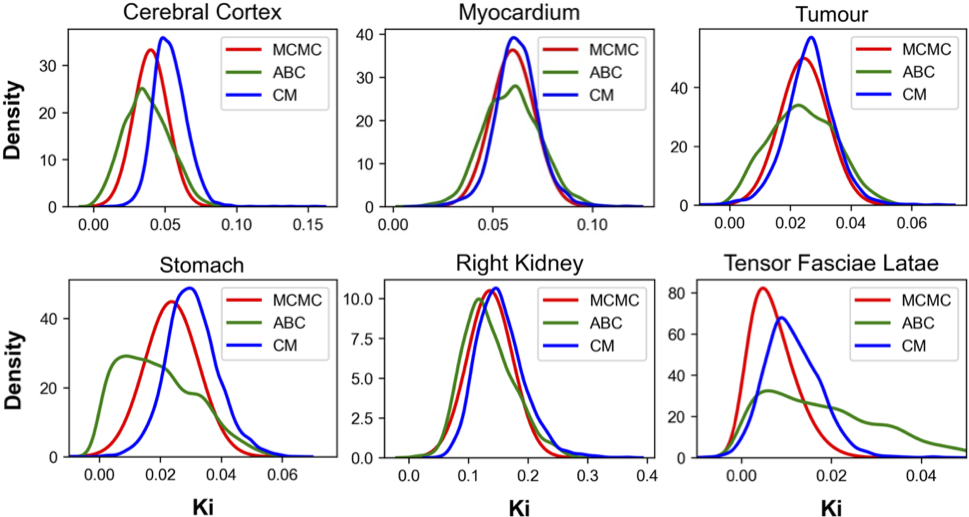}
    \caption{\textbf{Posterior predictions of $K_i$ in selected regions.} Six representative voxels are marked on the SUV image in Fig. \ref{fig:tb-parametric-imaging-treatment} by green dots in the left frontal cortex, myocardium, tumor, stomach, left kidney, and tensor fasciae latae. Posterior $K_i$ distributions predicted by MCMC, ABC and CM for these voxels are shown in the density plots.
    }
    \label{fig: Voxel-wise Posterior Distributions in Clinical Tissues}
\end{figure}

CM derived $K_i$ posteriors from six representative voxels are selected and compared to posteriors predicted by MCMC and ABC, as shown in Fig. \ref{fig: Voxel-wise Posterior Distributions in Clinical Tissues}.
For voxels from cerebral cortex, myocardium and the liver tumor, CM posteriors almost coincide with the MCMC reference, whereas ABC exhibits broader tails, suggesting an overestimation of uncertainty.  
In the stomach voxel CM matches MCMC, while the ABC posterior is right skewed and its longer right tail yields a smaller posterior mean.
In the right kidney, ABC and CM align with MCMC.
In the tensor fasciae latae (a muscle that attaches to the iliac crest of the pelvis), the CM posterior peaks at a slightly higher $K_i$ value than the MCMC posterior exhibiting a comparable but marginally broader spread, while the ABC posterior shares the MCMC peak location yet displays a markedly elongated right tail.

\begin{figure}[htp]
    \centering
    \includegraphics[width=0.78\columnwidth]{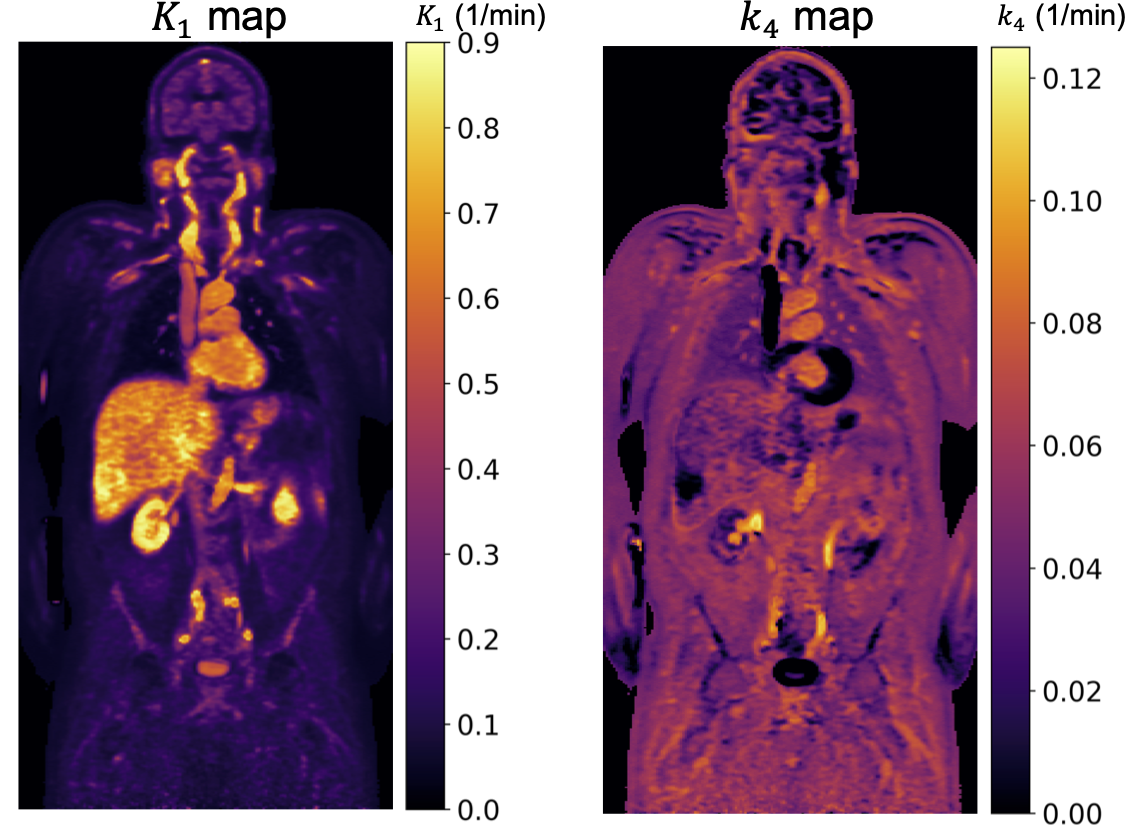}
    \caption{\textbf{Voxel-wise estimation of kinetic micro-parameters.} Total-body posterior mean maps for $K_1$ (delivery) and $k_4$ (clearance) generated by the CM.
    }
    \label{fig: P001 TB K1 k4 imaging}
\end{figure}

Fig. \ref{fig: P001 TB K1 k4 imaging} shows voxel-wise posterior mean maps for micro-kinetic parameters $K_1$ and $k_4$. The estimated values align with the established physiological kinetics. The $K_1$ map specifically shows high-perfusion regions, exhibiting maximal intensity in the kidneys, myocardium, and liver. In the $k_4$ map, the model distinguishes metabolic phenotypes. The hepatic tumour, myocardium, and cerebral gray matter exhibit negligible clearance ($k_4 \approx 0$) consistent with high FDG trapping, whereas the healthy liver displays non-zero dephosphorylation rates.

\subsection{Voxel-wise Model Selection}\label{section: model selection}
We used the posterior samples produced by the CM to create a voxel-wise binary parametric image of kinetic reversibility based on the efflux rate $k_4$. 
To generate this image, we assigned a 0 or 1 to each voxel indicating conformance with the irreversible ($k_4=0$) and reversible ($k_4>0$) two compartment models, respectively, based on the posterior probability that the efflux rate $k_4$ for the corresponding voxel is statistically distinguishable from zero. Specifically, for each voxel we examined the 95\% CrI of the posterior for $k_4$. If the interval included zero or if zero was on the right side of the CrI, the voxel was labelled irreversible. If the entire interval lay above zero, we classified the voxel as reversible.
Clinically, this map helps distinguish tissues in which tracer becomes metabolically trapped, such as cortical brain, myocardium and malignant tumours, from those where tracer exchanges reversibly with the free tissue compartment. 
Unlike point-estimation methods such as Patlak or nonlinear least squares, CM outputs a full posterior that can support model selection.

\begin{figure}[htp]
    \centering
    \includegraphics[width=\columnwidth]{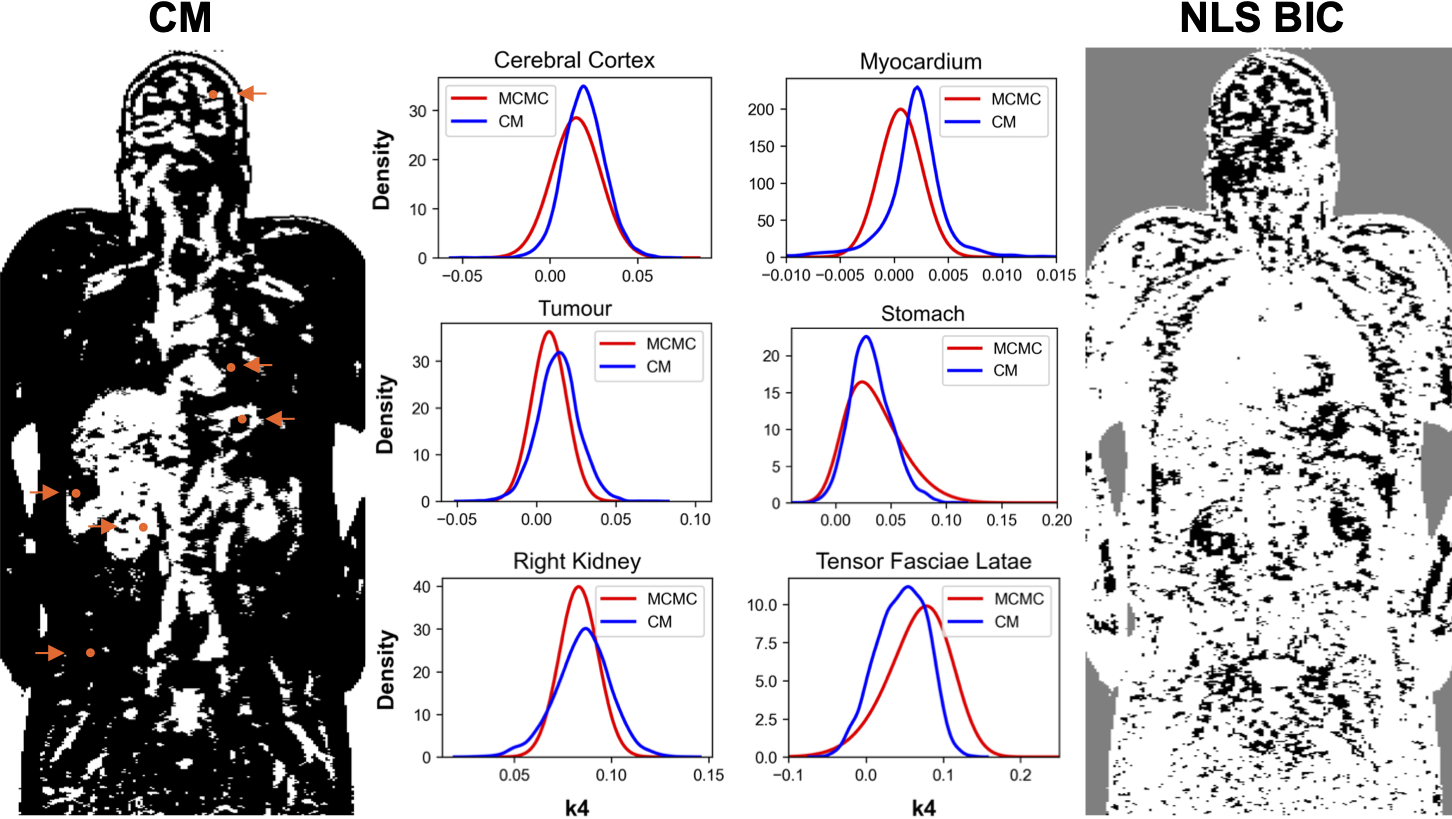}
    \caption{
    \textbf{Comparison of voxel-wise model selection.} 
    A binary mask labels voxels with a high posterior probability of irreversible efflux ($k_4=0$ black, $k_4 \neq 0$ white). Left: CM-derived model selection map (based on posterior $k_4$ support). Middle: Posterior $k_4$ distributions from CM (blue) and MCMC (red) are shown for six example voxels.
    Right: NLS-derived model selection map (based on BIC).
    }
    \label{fig: TB-model-selection-imaging}
\end{figure}

In Fig. \ref{fig: TB-model-selection-imaging} (left panel), brain, myocardium and skeletal muscle (indicated by the tensor fasciae latae) appear to be predominantly irreversible.
On the contrary, healthy liver and renal cortex are classified as reversible.
In particular, most liver voxels are classified as reversible, while the tumour in the inferior right lobe (previously highlighted in the $K_i$ map) shows a contiguous cluster of irreversible voxels.
Posteriors for six representative voxels illustrate the classification process (Fig. \ref{fig: TB-model-selection-imaging} middle panel). In cerebral cortex and myocardium the posteriors of $k_4$ predicted by CM are centred on the positive side of zero but the $95\%$ CrI contained the origin, hence both voxels are labelled as irreversible. 
The tumour voxel in the inferior right liver shows a similar slightly left-shifted probability density whose interval still contains zero, and is classified as irreversible.
By contrast, the voxel in the stomach and in the right kidney have posteriors that sit on the positive side of the axis, placing zero outside their $95\%$ CrIs, thus classifying them as reversible.
Finally, the tensor fasciae latae voxel exhibits a $95\%$ CrI containing zero, leading to an irreversible classification. 
Across all six voxels, the CM posteriors broadly mirror the distributions obtained by MCMC.

We compared our posterior-based model selection against standard frequentist approach. We calculated the Bayesian Information Criterion (BIC) for competing reversible and irreversible models using voxel-wise NLS estimator. As shown in Fig. \ref{fig: TB-model-selection-imaging} right panel, while BIC yields similar classifications in high-count organs (e.g., myocardium), it demonstrates instability in heterogeneous lesions. 
Specifically, NLS-BIC and CM showed inconsistencies in liver tumors. CM identified irreversible uptake kinetics in tumors ($k_4 \approx0$), and according to literature studies, the uptake of FDG by these tumors is more likely to be irreversible \cite{karakatsanis2015generalized,koukouraki2006comparison,okazumi1992evaluation}.
This underscores that direct inspection of the posterior CrI is not only more computationally efficient (avoiding multi-model fitting) but also reliable in low-SNR regions of voxel-wise TB-PET.

\section{Discussion}
% Major
% highlight the novelty of the method which is a fast and reliable generative AI-based and is able to produce total body parametric imaging of kinetic parameters with tens of millions ish number of voxels.
% highlight the speed and accuracy of the proposed CM in making parametric imaging and compare it to traditional methods like ABC and MCMC.
% compare with baselines.
% compare to previous works.
% discuss the rationale of the results presented in total body parametric imaging and in total body model selection imaging
% Minor
% need to discuss why we can compute Ki from K1, k2, and k3 rather than having a another new parameter for CM to predict.
This study describes a computational framework that produces fully Bayesian voxel-wise kinetic parameter maps, including full posterior distributions, across an entire TB-PET volume. By recasting posterior sampling as a conditional consistency mapping problem, the proposed CM reduces the hundreds to thousands of reverse diffusion steps required by DDPM and SBD into a single forward pass followed by a brief, fixed refinement schedule. 
This idea draws on the theoretical foundations of consistency models in computer vision \cite{song2023consistency} and extends them to medical imaging inverse problems at a scale that has not been reported previously in PET.
This method is a fast and reliable alternative to classical Bayesian tools such as ABC \cite{fan2021pet} and MCMC \cite{gelman2013bayesian} whose computational demands have so far confined rigorous posterior estimation to a limited number of regions of interest.
A systematic comparison with the classical approaches shows that CM preserves statistical accuracy but is about five orders of magnitude faster than MCMC. Across 100 synthetic test samples, posterior means for the micro-kinetic parameters deviated by $<5$\% from MCMC estimates, whereas ABC produced broader tails and required tens of millions of forward simulations per voxel. 
Processing 1 slice of TB-PET data (approx. 83,000 voxels) with CM completed in 1 min (using 4 NVIDIA A100 GPUs), compared with 2 min for GPU-accelerated ABC (1 NVIDIA A100 GPU) and over 1 month for parallel MCMC on a 128-core CPU cluster. Additionally, the CM completed processing the entire TB-PET data (4.4 million voxels) within 1.25 hours on the 4-GPU setup.
Our study thus paves the way for routine Bayesian whole-body parametric imaging with uncertainty estimation.

We also demonstrated that CM outperforms modern deep-learning baseline methods. 
DDPM and SBD match the prediction accuracy of CM but require hundreds to thousands of reverse diffusion steps, which significantly slows down the sampling. 
CVAE systematically underestimates variance mainly because the objective function minimizes the forward Kullback–Leibler term, which harshly penalizes allocating probability mass to regions where the true posterior is low \cite{alemi2018fixing,dieng2017variational}.
GAN suffers from biased posterior means and underestimated CrIs, since the generator need only fool the discriminator on the regions it covers; this incomplete coverage whether by missing modes or by over-concentrating on part of a single mode shifts the empirical mean and leads to narrow CrIs \cite{arora2017gans,lucic2018gans}.

When positioned in the context of the wider literature, CM addresses several gaps. Djebra et al. recently proposed a DDPM for posterior prediction in dynamic PET \cite{djebra2025bayesian}, but their experiments were limited to a few thousand simulated TACs, a limited number of regions of interest and the method has not been demonstrated on total body PET data. 
Other learning-based methods, such as ParaPET \cite{vashistha2024parapet}, SN-Patlak \cite{gu2025self} and related ROI-level CVAE frameworks \cite{liu2023posterior}, deliver either single-value point estimates or parametric images at much lower spatial resolution. 
Conventional Patlak total body methods \cite{rahmim2019dynamic,li2025total} assume irreversible kinetics and cannot distinguish reversible from irreversible trapping, whereas CM generates posterior distributions for all micro-parameters and an explicit voxel-wise model selection map, thereby adding physiological interpretability lacking in these algorithms.

The biological plausibility of the CM outputs further supports the clinical relevance. 
The $K_{i}$ images reproduce canonical FDG uptake patterns with high net influx values in cerebral cortex, myocardium and renal cortex, intermediate values in liver and skeletal muscle, and low values in lung parenchyma, consistent with large-cohort reference values for dynamic whole-body FDG scans \cite{dias2022normal,liu2021kinetic}. 
Moreover, the model selection map classifies brain, myocardium and tumour voxels as effectively irreversible, reflecting negligible de-phosphorylation of FDG-6-P, while kidney cortex and much of the hepatic parenchyma are labelled as reversible, in agreement with known renal FDG wash-out and hepatic glucose-6-phosphatase activity reported in the literature \cite{dias2022normal}. 

The clinical utility of uncertainty estimation enables more robust decision-making, particularly in the high-noise environments inherent to voxel-level total-body PET.
In these noisy conditions, the inverse problem of kinetic modeling becomes ill-posed. 
The likelihood surface is often flat or complex, causing standard point estimators (e.g., NLS) to exhibit high variance or converge to local minima. In contrast, the CM provides the full posterior distribution, capturing the uncertainty of model parameters due to noisy measurements. We demonstrate the practical value of this via the voxel-wise model selection (Fig. \ref{fig: TB-model-selection-imaging}), where posterior distributions differentiate reversible from irreversible kinetics. A further plausible clinical translation is the use of posterior variance as a voxel-wise reliability index, enabling clinicians to distinguish high-confidence metabolic features from noise artifacts that require further investigation.

Two systematic discrepancies emerge when contrasting the Patlak derived and CM derived $K_i$ maps. First, Patlak slopes are uniformly attenuated relative to CM estimates because the measured tissue curves are not corrected for vascular spill‑in. Therefore, the resulting Patlak slope corresponds to $(1\!-\!V_b)\,K_i$, so even modest blood‑volume fractions (3--5\,\% in most tissues, higher in highly perfused organs) translate into lower $K_i$ values. 
Second, the spatial correspondence between Patlak and CM maps is imperfect. Patlak relies on the assumption of irreversible tracer trapping (\(k_4=0\)) to fit a straight line at mid-late time points \cite{zuo2020multiphase}.  
This assumption is violated in organs with predominantly reversible kinetics, and elevated noise in low‑count voxels further destabilises the fit, thus introducing artefacts and regional bias \cite{karakatsanis2015generalized}.
While NLS is often considered a standard estimator for the full two tissue compartment models, it is known to exhibit bias, particularly when inequality constraints are applied to model parameters \cite{escobar1986bias}. In contrast, MCMC provides an asymptotically unbiased estimate of the posterior distribution. By training the CM to approximate the MCMC posterior, our method aims to inherit this unbiased property while bypassing the computational cost that renders MCMC impractical for total-body PET.

Many implementations of model selection metrics, such as AIC, typically rely on the residual sum of squares (RSS) derived from fitting competing models \cite{wang2018dynamic,wang2022total,zuo2020multiparametric}. However, using this RSS-based formula implicitly asserts that the noise follows a Gaussian distribution—an assumption that is often inaccurate for reconstructed images. Furthermore, these methods rely on point estimates (MLEs) and asymptotic assumptions that are unstable in high-noise, low-count voxel data \cite{cavanaugh2019akaike}. 
In contrast, our proposed framework enables model selection via a direct probability statement in a single inference pass. Being likelihood-free, it avoids the need for valid noise assumptions or arbitrary complexity penalties. Instead, we inspect the 95\% CrI of $k_4$. If the CrI for $k_4$ includes zero, the data is deemed consistent with irreversible kinetics. This allows for probabilistic model selection that naturally integrates parameter uncertainty.

Although the present network was trained solely on FDG data, its modular encoder–decoder architecture makes adaptation to other tracers (e.g., [\textsuperscript{18}F]FET, [\textsuperscript{18}F]FMISO) or alternative kinetic structures (e.g., one-tissue or dual-input models) straightforward. Parameter efficient techniques such as adapters and low‐rank (LoRA) fine‐tuning \cite{houlsby2019parameter,hu2022lora} can graft tracer‐specific or model-specific kinetics onto the existing backbone with only a few thousands training samples. This is yet to be demonstrated but will be investigated in the next stage of this research.

The selection of the CM was motivated by the specific topological properties of kinetic analysis. Unlike high-resolution image generation where the data manifold is highly curved, often necessitating trajectory straightening techniques like Rectified Flow \cite{liu2022flow} or Flow Matching \cite{lipman2022flow}, voxel-wise 1D kinetic estimation involves a topologically simpler manifold where the direct mapping learned by CM is highly effective. However, we acknowledge that the field is rapidly evolving. Recent innovations such as Improved Mean Flows \cite{geng2025improved} and Continuous-Time Consistency Models \cite{lu2024simplifying} offer enhanced training stability and prediction accuracy. Future work will explore extending the current framework with these advanced methods.

Several limitations of our study warrant acknowledgment.
First, the training dataset covered a limited range of noise levels based on eight subjects. Acquisition settings with substantially higher noise (e.g., ultra-low-dose protocols, motion corruption) may exceed the model’s learned range, requiring retraining or fine-tuning with synthesized high-noise samples, or prior denoising.
Second, regarding generalizability across different scanners and reconstruction protocols, we note that while the current model is robust to a range of noise magnitude (due to the wide noise range $\lambda \in [3, 7]$ used in training), it assumes a specific noise texture and temporal framing (i.e., sampling schedule). A 'universal' model capable of handling arbitrary framings or reconstruction kernels is beyond the scope of this work. However, the proposed framework serves as a foundational model and adaptation to different scanner geometries or reconstruction-induced noise correlations can be achieved via parameter-efficient fine-tuning (e.g., LoRA adapters \cite{hu2022lora}).
Third, regarding tracer arrival times, our CM framework is inherently capable of handling variable delays, having been trained on offsets ($t_0$) ranging from 0 to 1.25 minutes. However, the current clinical application utilizes a single global input function derived from the aorta. Consequently, while the CM is robust to temporal misalignment (as shown in Fig. \ref{fig: Params error vs tracer arrival time}), it does not explicitly model the specific physiological delay and dispersion differences between organs, as this would require organ-specific AIFs. Implementing such an extension by deriving local input functions from the imaging data remains an important direction for future work.
Finally, this study applied the two tissue compartment model as a universal kinetic structure, while specific regions (e.g., lungs or bone) may benefit from alternative models. 
Future work will extend this framework to perform Bayesian model selection together with posterior estimation of kinetic parameters.

\section{Conclusion}
We present PET-TRACER, a consistency model framework that generates voxel-wise kinetic parameter maps and model selection maps from dynamic TB-PET data. The method matches MCMC accuracy, performs competitively with recent deep-learning approaches, and greatly reduces computation time. It is particularly suited for studying kinetic heterogeneity in multi-organ TB-PET and systems physiology. PET-TRACER is available on GitHub: \url{https://github.com/yundumbledore/PET-TRACER}.

\section*{Acknowledgment}
The authors thank Paul Roach, Dale Bailey, Sally Ayesa and Liz Bailey for their advice on protocol design and the nuclear medicine staff at Royal North Shore Hospital for performing the TB-PET studies. We also acknowledge the facilities and the scientific and technical assistance of Sydney Imaging, a core research facility at The University of Sydney, and the National Imaging Facility, a National Collaborative Research Infrastructure Strategy (NCRIS) Capability.

\bibliographystyle{IEEEtran}
\bibliography{bib}

\end{document}